\RequirePackage{ifpdf}
\documentclass[11pt,a4paper]{article}
\usepackage{jcappub}

\usepackage{epsfig}
\usepackage{comment}
\usepackage{amsmath}

\def\vo{\mathcal{V}}
\def\mc{\mathcal}
\def\ba{\begin{eqnarray}}
\def\ea{\end{eqnarray}}
\def\be{\begin{equation}}
\def\ee{\end{equation}}

\makeatletter
\def\x@arrow{\DOTSB\Relbar}
\def\xlongequalsignfill@{\arrowfill@\x@arrow\Relbar\x@arrow}
\newcommand{\xlongequal}[2]{%
    \ext@arrow 0099\xlongequalsignfill@{#1}{#2}}
\makeatother

\newcommand{\roughly}[1]{\mathrel{\raise.3ex\hbox{$#1$\kern-0.85em
\lower1ex\hbox{$\sim$}}}}

\newcommand{\gsim}{\roughly>}
\def\nott#1{\setbox0=\hbox{$#1$}                
   \dimen0=\wd0                                 
   \setbox1=\hbox{/} \dimen1=\wd1               
   \ifdim\dimen0>\dimen1                        
      \rlap{\hbox to \dimen0{\hfil/\hfil}}      
      #1                                        
   \else                                        
      \rlap{\hbox to \dimen1{\hfil$#1$\hfil}}   
      /                                         
   \fi}                                         %

\def\endignore{}
\def\ignore #1\endignore{}
\def\be{\begin{equation}}
\def\beq\begin{equation}
\def\ee{\end{equation}}
\def\bea{\begin{eqnarray}}
\def\eea{\end{eqnarray}}

\def\eqa{\begin{eqnarray}}
\def\eeqa{\end{eqnarray}}
\def\eq{\begin{equation}}
\def\eeq{\end{equation}}

\def\pref#1{(\ref{#1})}

\def\pref#1{(\ref{#1})}
\def\be{\begin{equation}}
\def\ee{\end{equation}}

\def\beq{\begin{equation}}
\def\eeq{\end{equation}}
\def\beqa{\begin{eqnarray}}
\def\eeqa{\end{eqnarray}}

\def\ssA{{\scriptscriptstyle A}}
\def\ssB{{\scriptscriptstyle B}}

\def\ssM{{\scriptscriptstyle M}}
\def\ssN{{\scriptscriptstyle N}}

\def\WL{{\scriptscriptstyle WL}}
\def\UV{{\scriptscriptstyle UV}}
\def\IR{{\scriptscriptstyle IR}}

\newcommand{\bmat}{\left(\begin{array}}
\newcommand{\emat}{\end{array}\right)}

\def\-{\hphantom{-}}

\def\s2{\frac{1}{2}}

\def\IF{\relax{\rm I\kern-.18em F}}
\def\II{\relax{\rm I\kern-.18em I}}
\def\IP{\relax{\rm I\kern-.18em P}}
\def\IC{\relax{\rm I\kern-.48em C}}
\def\IIR{\relax{\rm I\kern-.18em R}}
\def\IK{\relax{\rm I\kern-.20em K}}
\def\IM{\relax{\rm I\kern-.25em M}}

\def\y2{Y_{\ssM\ssN} Y^{\ssM\ssN}}
\def\Riem2{R_{\ssA\ssB\ssM\ssN} R^{\ssA\ssB\ssM\ssN}}
\def\Ricci2{R_{\ssM\ssN} R^{\ssM\ssN}}

\def\f2{F^{a}_{\ssM\ssN} F^{\ssM\ssN}_a}

\def\Asl{\hbox{/\kern-.7500em\it A}}
\def\dsl{\hbox{/\kern-.5500em$\partial$}}
\def\pxpsl{\hbox{/\kern-.5600em$p$}}
\def\Dsl{\,\raise.15ex\hbox{/}\mkern-13.5mu D}

\def \one{\relax{\rm 1\kern-.26em I}}

\def\({\left(}
\def\){\right)}

\title{String Inflation After Planck 2013}

\author[a,b]{C.P.~Burgess,}
\author[c,d,e]{M.~Cicoli,}
\author[e,f]{F.~Quevedo}

\affiliation[a]{Department of Physics \& Astronomy, McMaster University, Hamilton ON, Canada}
\affiliation[b]{Perimeter Institute for Theoretical Physics, Waterloo ON, Canada}
\emailAdd{cburgess@perimeterinstitute.ca}
\affiliation[c]{Dipartimento di Fisica e Astronomia, Universit\`a di Bologna,\\ $\qquad$ via Irnerio 46, 40126 Bologna, Italy}
\affiliation[d]{INFN, Sezione di Bologna, Italy}
\affiliation[e]{Abdus Salam ICTP, Strada Costiera 11, Trieste 34014, Italy}
\emailAdd{mcicoli@ictp.it}
\affiliation[f]{DAMTP, University of Cambridge, Wilberforce Road,  Cambridge, CB3 0WA, UK}
\emailAdd{F.Quevedo@damtp.cam.ac.uk}

\abstract{We briefly summarize the impact of the recent Planck measurements for string inflationary models, and outline what might be expected to be learned in the near future from the expected improvement in sensitivity to the primordial tensor-to-scalar ratio. We comment on whether these models provide sufficient added value to compensate for their complexity, and ask how they fare in the face of the new constraints on non-gaussianity and dark radiation. We argue that as a group the predictions made before Planck agree well with what has been seen, and draw conclusions from this about what is likely to mean as sensitivity to primordial gravitational waves improves.}

\begin{document}
\maketitle

\section{Introduction}

The problem with making predictions is that people test them. This is a relatively unfamiliar problem for string theorists which with luck may be beginning to change due to the observational access to the very early universe made possible by the development of precision cosmology. Most recently has come the release of yet more precise cosmological results from the Planck satellite \cite{Planck}, which seems to confirm the standard concordance Hot Big Bang cosmology with exquisite precision \cite{PlanckParam}. And more is potentially in store, including significant increases in the sensitivity to primordial gravitational waves.

This raises a (widely appreciated) opportunity. The success of concordance cosmology is contingent on the existence of primordial fluctuations having specific properties, and a physical understanding of these properties requires understanding the much-earlier epochs from which they emerge. But these epochs probe temperatures and energies much higher than those to which we have access elsewhere, and so open an observational window onto physics we might otherwise not have been able to probe.

So given the flood of new information, it is timely to step back and assess what it is telling us about fundamental physics, and how the various theoretical models are doing so far. In this paper we provide our own preliminary assessment, in particular for string inflationary models. Before doing so, however, given the complexity of these models it is worth first asking why they are worth scrutinizing in detail at all. After all, if the data is perfectly consistent with much simpler models, Occam's razor suggests we should leave it at that. (Those who need no convincing on this score should jump ahead to \S\ref{sec:scorecard}.)

\subsection{Why consider such complicated models?}

Broadly speaking there are two ways of thinking about the kind of theoretical description that should be sought to explain the data. One approach is to find as simple a model as possible that is consistent with observations, with more complicated models only to be considered to the extent that they are motivated by unexplained features of the data \cite{Slava}. This approach is the one traditionally used in cosmology.

A second approach, more common amongst particle physicists, takes the point of view that what is being observed is only the low-energy limit of something more fundamental. This point of view is driven both by past experience with particle theories such as the Standard Model, and by the knowledge that gravitational theories are non-renormalizable and the only known way to quantify the uncertainties in predictions with such theories is within the context of a low-energy expansion.\footnote{This is particularly true in situations where quantum effects are believed to be significant (such as the seeds for primordial fluctuations during inflation).} Experience with low-energy limits throughout physics shows that they are often messy and not so simple as might otherwise be desired.  This point of view leads one to ask what features are {\em generic} to the degrees of freedom that could be at play at the relevant energies. Of course this doesn't mean dropping simplicity in favour of complexity for its own sake. But it does encourage thinking more broadly about all the degrees of freedom that might play a role, and thinking about the circumstances under which simplicity might nevertheless emerge in any case.

For something like cosmology this second approach might seem pretty hopeless, given we have very limited information about what physics is like at the energies relevant to primordial fluctuations. But experience teaches that Nature tends to hide most of the details of short-distance physics from longer-distance observations. Ultimately this is why progress in science is possible; each hierarchy of scale can be understood largely on its own terms. But there are usually a few long-distance issues that {\em do} hinge on more-microscopic details, and when these are found they can be tremendously informative about the mechanics of much shorter distances.

\subsection{Understanding acceleration}

The properties of primordial fluctuations appear to be among these high-energy-sensitive issues. So what can these cosmological observations really tell us about all the minutae of short-distance physics? The emerging picture is one of striking simplicity. The observed scalar fluctuations are Gaussian \cite{PlanckNG} and have a scalar spectral index that is close to (but not exactly the same as) the scale-invariant value: $n_s = 1$. There is no evidence yet for tensor fluctuations (primordial gravitational waves) \cite{PlanckInf}.

These properties strongly resemble what would emerge from a much earlier epoch of accelerated universal expansion; which in principle could be either an `inflationary' acceleration between epochs of slower expansion \cite{Inflation}, or a `bouncing' acceleration that occurs between epochs of universal contraction and expansion \cite{Bounces}. But in either case successful descriptions require the underlying physics to have unusual features, making it useful to explore how they can be formulated within a more complete formulation of quantum gravity, such as string theory.

On the bounce side the unusual feature is the bounce itself. General arguments make it difficult to reverse a gravitational collapse within stable systems (such as those satisfying the null energy condition). So proponents of these models rightly complicate them by embedding into string theory \cite{Turok:2004gb}, in ways designed to see if this can be possible. This usually involves finding a formulation for which strong gravitational fields apply, such as string theory. We do not explore this option further, but see \cite{PlanckCyclic} and \cite{Issues} for recent discussions of the advantages and challenges in doing so.

But bouncing cosmologies are not the only ones that benefit from having a more fundamental formulation. Inflationary models also have some unsatisfactory features \cite{SteinhardtString} on which a more fundamental formulation can shed light, of which we list four:
\begin{itemize}
\item {\em Abnormally flat potentials (the $\eta$-problem):} The first is the requirement to have scalar fields that are so light that the effects of their mass does not compete with the extremely weak gravitational effects due to universal expansion. It is rare to find theories with a scalar mass satisfying $\mu \ll M$, where $M$ is the basic fundamental scale of the theory. (In the context of the Higgs scalar, this is the electro-weak hierarchy problem.) But it is even more unusual to find a scalar whose mass satisfies $\mu \ll M^2/M_p$, as is required of inflationary models. Controlling this requires being able to control quantum interactions of gravitational strength, as is at present only possible in a precise way in string theory.
\item {\em Trans-Planckian field motion:} Second, some predictions (like observable primordial gravitational waves) may require fields to travel over Planckian distances in field space. Because predictions in gravity rely inherently on the existence of a low-energy expansion relative to a mass scale that is at most the Planck scale, control over all approximations can be tricky once fields start to roll out to Planckian values. Large fields need not imply large energies, which is why trans-Planckian field values are not simply crazy to entertain, but so far only string theory provides a framework in which the detailed implications of such large fields can be tracked in a precise way.
\item {\em Eternal inflation:} Third, inflation can go viral, inasmuch as it can be hard to end once it begins. Even if it is stamped out in one area of the universe it need not end everywhere at the same time. And the regions where it occurs tend to expand so quickly that they dominate the later stages of the universe. This feature is particularly vexing in models with a complicated potential landscape wherein fields can find a variety of local minima, many of which may inflate. In such theories it can be difficult to know how to predict unambiguously what observers in the distant future will see.\footnote{See \cite{Bubbles}, however, for an attempt to find observable consequences within the eternal inflation picture.}
\item {\em Initial conditions:} Many inflationary scenarios only work if the fields are initially very homogeneous and/or start with precise initial positions and velocities. Any physical understanding of this is likely to be pushed back to still-earlier epochs with ever-higher energies, for which a more complete formulation (like string theory) is likely required.
\end{itemize}

All of these questions motivate asking how inflationary model-building might fit within more complicated theories like string theory, and the second of them is likely to have some impact on how likely we are to find primordial gravitational waves. So far little progress has been made with the last two, and so there is little we can add about these at this point beyond making the following remarks.

The issue of eternal inflation is a thorny one, that is closely related to the `landscape' issue. Fundamental theories like string theory have a large number of degrees of freedom compared with ordinary field theories, and are expected to have a complicated potential energy that admits an enormous number of stable or quasi-stable `vacuum' solutions. Because the theory is generally covariant, energy is not available as a criterion for choosing among them to identify the `real' vacuum. The landscape issue asks how one should make predictions in a theory with a large number of stable (or quasi-stable) vacuum solutions, without knowing in detail where we live among them. Eternal inflation makes this problem worse by exponentially expanding the volume of any region containing a vacuum whose equation of state is inflationary.

Although we agree this is an important problem of principle that needs study, we believe there are two reasons for not yet using this to discard inflationary models. First, it has not yet been an obstacle to making practical predictions in the less ambitious problem of inferring what cosmic fluctuations look like in a part of the universe in which an inflationary period has just ended. This is because the inflationary framework does not intrinsically require control over the physics of strong gravity, such as for a bounce. Indeed calculations with the simplest models are what the Planck collaboration compares to when finding such impressive agreement with their observations \cite{PlanckInf}.

Second, the problem of eternal inflation is not really a problem specific to inflationary models for the CMB. Instead, it is potentially a problem for any kind of cosmology based on a landscape. This is because it is insufficient to successfully engineer a desirable cosmology in one corner of the landscape --- be it an inflationary model or a bounce or something else. One must then always check that eternal inflation doesn't also appear {\em anywhere} else in any of the other vacua throughout the entire theory \cite{EternalLandscape}. To the extent that eternal inflation is a fundamental challenge, it is equally a challenge to all who use landscape-prone physics (like any quantum gravity, so far as is known) as any part of their story.

\section{A string-inflationary score card}
\label{sec:scorecard}

So how do models of string inflation do when compared with the Planck data? Naively, one might think not well, since string models generically predict a complicated, often multi-field, evolution whereas observations seem clearly to favour simple single-field models \cite{Single-field}. This turns out to be naive because the observational consequences of multi-field evolution are often well-described by an effective single-field model. Sometimes this `emergent' single-field description arises because only one field is light enough to be rolling at the relevant epoch, but more often it occurs because the multiple fields do not do anything complicated at the epoch of horizon exit.\footnote{More precisely, they move along a target-space geodesic.}

Let us start with a brief discussion of the main proposals for string inflation \cite{StringCosmologyRevs,CQreview}. For definiteness we restrict ourselves to models which try to address modulus stabilisation since this is known to be a crucial issue when predicting the dynamics of the inflaton field.\footnote{See, however, also \cite{Mflation} for an interesting string inspired inflationary model without modulus stabilisation.} The known inflationary models split into two main classes according to the nature of the inflaton field which can be either an open or a closed string modulus. For each of these classes we focus on those cases where the effective single-field scalar potential can be either explicitly computed or be well motivated from string theory. We also briefly summarize some of their main features and phenomenological predictions.

\subsection{Open string inflationary models}

Let us briefly outline the main features of models where the inflaton is an open string mode:

\begin{enumerate}
\item{} {\it $D3/\overline{D3}$ Brane Inflation}

The first string models with calculable modulus stabilisation \cite{KKLMMT} invoked the mechanism of brane-antibrane inflation \cite{BBbarInf, Dvali:2001fw} for which the inflaton, $\phi$, describes the distance between a brane and an anti-brane. The single-field potential relevant to the initial models had the form:
\be
V \simeq V_0\left(1-\frac{\mu^4}{\phi^4}\right)\,,
\label{VKKLMMT}
\ee
where $V_0 = 2 T_3 h^4$ with $T_3$ the tension of a D3-brane and $h$ the warp factor, while $\mu \propto V_0^{1/4}$. This leads to the following predictions for $50\leq N_e\leq 60$: $0.966\leq n_s\leq 0.972$, $r\leq 10^{-5}$ and vanishing non-Gaussianities. Matching the Planck prediction for the running of the spectral index within $2 \sigma$ requires $10^{-29}\leq \mu/M_p \leq 0.1$, which in turn
gives an inflationary scale of order $10^6\,{\rm GeV}\leq M_{\rm inf}\leq 10^{15}\,{\rm GeV}$ \cite{KKLMMTPlanck}.

\item{} {\it Inflection Point Inflation}

A criticism of brane-motion models is that the scalar potential (\ref{VKKLMMT}) receives corrections, whose form must be computed in detail to check if it is possible to obtain a sufficiently flat potential. This is a special case of the $\eta$-problem, which expresses how difficult it is to get scalars with masses satisfying $m^2 \ll H^2$, as is required of an inflaton. For example, contributions from `Planck slop' of the form:
\be
\Delta V  \propto \frac{V_0\,\phi^2}{M_P^2}\,,
\label{LargeEta}
\ee
produce unacceptably large corrections to the $\phi$ mass.

More precise calculations of the potential have been done for the KKLMMT model to show that the tuning can be explicitly achieved, leading to an inflection-point inflationary model whose effective single-field potential has the form \cite{delicate}:
\be
V\simeq V_0 \left(1+\lambda_1\phi+\frac{\lambda_3}{3}\,\phi^3 \right)\,.
\ee
Given that the potential is flat only in a small region around an inflection point where $\eta=0$,
in order to obtain large numbers of $e$-foldings, $N_e\propto 1/\sqrt{\epsilon}$,
the slow-roll parameter $\epsilon$ needs to be much smaller than $|\eta|$.
Hence the fact that one obtains a red or blue spectral index, $n_s -1 \simeq 2 \eta$,
depends just on the sign of $\eta$ at horizon exit. This sign, in turn, depends on
the total number of $e$-foldings $N_e^{\rm tot}$, because it depends on whether horizon exit takes place above or below the inflection point on the scalar potential. For instance, if one requires just $N_e^{\rm tot}\simeq 60$, then $\eta>0$
at horizon exit, resulting in a blue spectrum; $n_s > 1$. On the other hand, if $N_e^{\rm tot}\simeq 120$,
one has about $60$ $e$-foldings above the inflection point and $60$ $e$-foldings below,
giving a scale-invariant spectrum; $n_s \simeq 1$. For more than 120 $e$-foldings the spectrum is red and so $n_s < 1$.

This correlation of $n_s > 1$ for fewer $e$-foldings of inflation can be a worry if inflation is regarded as a process that occurs randomly as the inflaton occasionally encounters inflection points within a complicated potential. This is because simulations of this process reveal inflationary epochs with fewer $e$-foldings to be more probable than those with many $e$-foldings, with the probability of $N_e$ $e$-foldings of inflation being proportional to $N_e^{-3}$ \cite{Random}. Consequently most of the random inflationary events that occur in this way would predict a blue spectrum.

Putting aside these statistical considerations, we focus on the special case where $N_e$ is large enough to produce a red spectrum, as observed. Taking for simplicity $\lambda_1=0$ the spectral index becomes $n_s-1\simeq -4/N_e$ (setting the end of inflation at $\phi_{\rm end}\to -\infty$)
implying that for $50\leq N_e\leq 60$ the spectral index lies in the low range $0.92 \leq n_s \leq 0.93$ \cite{lindewestphal}. This is clearly an effective small single-field inflationary model with $r\leq 10^{-6}$ and negligible non-Gaussianities. The running of the spectral index becomes:
\be
\alpha_s \equiv \frac{d n_s}{d \ln k} \simeq - \frac{4}{N_e^2} \,,
\ee
giving the prediction $-0.0012\leq \alpha_s\leq -0.0008$. The inflationary scale is set by $V_0^{1/4}$ and typically turns out to be of order the intermediate scale. For example, matching the COBE normalisation of the amplitude of density perturbations at $N_e\simeq 60$,
one finds $M_{\rm inf} \simeq 5\cdot 10^{12}$ GeV. Tensor modes are therefore not detectable in such a scenario.

\item{}
{\it DBI Inflation}

Dirac-Born-Infeld (DBI) models \cite{DBI} are a class of inflationary models that is more intrinsically stringy than most, arising when brane motion occurs through a strongly warped region within the extra dimensions. This in the sense that they do not involve a standard slow-roll dynamics but non-canonical kinetic terms and the presence of a speed limit in a warped space-time \cite{DBI}. Their dynamics is described by:
\be
{\cal L}_{\rm DBI}= -f(\phi)^{-1} \sqrt{1-2 f(\phi) g^{\mu\nu} \partial_\mu \phi \partial_\nu \phi} + f(\phi)^{-1} - V(\phi) \,,
\ee
where $f(\phi)\sim A/\phi^4$ is a warp factor.

This is one of the very few kinetic modifications of single-field models that can be sensibly understood within the low-energy approximation that is implicit in cosmological applications of gravity. This is because the DBI Lagrangian is one of a class of Lagrangians that depend in a non-polynomial way on $X \equiv \partial_\mu \phi \partial^\mu \phi$, but without also including equal numbers of higher derivatives of $\phi$. Usually this would be inconsistent for a low-energy effective field theory, but not so for the DBI action, which is protected by symmetries \cite{NLRSpacetime}.

Two types of scalar potential, $V(\phi)$, have been considered:
\be \label{DBIpots}
V_{\UV} \propto \phi^2 \quad \hbox{\cite{DBI}}
\qquad \hbox{or} \qquad
V_{\IR} \simeq V_0 - \frac{\beta}{2} \, H^2\phi^2 \quad \hbox{\cite{IRDBI}} \,,
\ee
corresponding to brane motion in two different kinds of warped regions. $V_\UV$ was used in the original formulation of DBI inflation, with a brane moving from the `UV' to the `IR' regime of the warped region. This is disfavoured by consistency arguments \cite{Chen} and experimental constrains. On the other hand, the second case is a small-field inflationary model whose predictions: $n_s-1\simeq - 4/N_e$, $\alpha_s\simeq -4/N_e^2$ and unobservably small $r$ --- for instance ref.~\cite{lidseyhuston} finds $r \leq 10^{-7}$ --- are similar to the case of inflection-point inflation.

Warping is strong in regions where $f(\phi)$ is large, and because the inflaton motion is then not of the slow-roll type, 
these models naturally give rise to potentially large equilateral non-Gaussianities whose observational consequences we describe below.

\item{} {\it Wilson Line Inflation}

The original proposal \cite{wilsonline} was based on the fact that Wilson lines are T-dual to branes intersecting at angles in which angles are dual to magnetic fluxes and brane separations to Wilson lines. Given that intersecting branes (before moduli stabilisation)
were good candidates for inflaton fields,
Wilson lines could also be used as candidates for inflatons in IIB models that can also include moduli stabilisation. Wilson lines are also present in heterotic models where brane separation moduli are not present. In the small field regime the potential is of the form:
\be
V_{\WL}=A-\frac{B}{\phi^2}\,,
\ee
with similar features as brane inflation.
Also a warped DBI version of Wilson line inflation was considered in \cite{warpedWL} in which it is possible to have large $r$ and large $f_{\rm NL}$ with a lower bound for $r$ that increases with the bounds on $f_{\rm NL}$ creating already a tension with current observations similar to DBI inflation.

\item{}
{\it D3-D7 Inflation}

In D3-D7 models the inflaton is the position of a D3-brane that moves relative to the position of higher-dimensional D7 branes \cite{D3D7}. This construction has the promise of providing inflationary examples with a supersymmetric
final state, inflation driven by D-terms, and a potentially interesting cosmic-string signature.\footnote{For models where the inflaton is the
distance between two fluxed D7-branes see \cite{FluxInfl}. The predictions of this model are
$n_s-1\simeq -1/N_e$, which for $N_e \simeq 60$ gives $n_s\simeq 0.983$, and unobservable tensor modes (see however \cite{Arthur} for 
ways to obtain $r \sim 10^{-3}$).
One problem is that the embedding of this model in compact constructions with stabilised moduli
yields an effective field theory which is only marginally under control.}
In the first approaches the potential was argued to have a logarithmic regime:
\be
V = V_0 + A \ln \phi - B \phi^2 + C \phi^4 \,,
\ee
with the prediction $n_s \simeq 0.98$ \cite{D3D7}. More detailed constructions allow a smaller $n_s$ \cite{HAACK}.
Ref.~\cite{K3T2} provides a careful calculation of the uplifted string potential for the D3-D7 inflationary system. Although a simple single-field description was not found, numerical integration did reveal inflation with the inflaton being a mixture of the brane position and a modulus describing the shape of the extra dimensions, with predictions ($n_s \simeq 0.96$ and $r \simeq 0$) resembling those of `racetrack inflation' described below.
\end{enumerate}

\subsection{Closed string inflationary models}

In this section we outline the main features of some models where the inflaton is a closed string mode.
These models split into two categories according to the nature of the inflaton field as an axion or
the volume of an internal four-cycle.

\subsubsection{Inflation using axions}

Let us start describing models where the inflaton is an axion field \cite{AxionReview}.
An attractive feature of these scenarios is the axionic shift symmetry
which forbids the presence of inflaton-dependent higher dimensional operators
which could give rise to dangerously large contributions to the slow-roll parameter $\eta$.
This is not enough to solve the $\eta$-problem since one has also to find a model
where the potential built from renormalisable operators is flat enough to drive inflation.
Here are some promising inflationary models within this category:

\begin{enumerate}
\item{} {\it Racetrack Inflation}

This was the first explicit model of inflation in string theory in which the scalar potential was explicitly computed including moduli stabilisation and the inflaton was a closed string mode \cite{Racetrack}. The explicit form of the potential is not informative since it is a complicated two-field inflationary model. The inflaton is mostly an axion field which is a component of a K\"ahler modulus.  It is hard to extract a simple, single-field effective potential during inflation, since the single-field inflationary trajectory was found numerically. Also though it was the first successful case of closed string moduli inflation, the strong fine tuning makes these models unattractive theoretically at the moment.
It corresponds to small field inflation, almost Gaussian with $n_s\simeq 0.96$. The inflationary scale
is of order $M_{\rm inf}\simeq 10^{14}$ GeV corresponding to a very small tensor-to-scalar ratio
$r\simeq \left(M_{\rm inf}/M_{\rm GUT}\right)^4 \simeq 10^{-8}$.

\item{}{\it N-flation}

In this scenario $N\gg 1$ axions conspire to provide successful inflationary conditions with a Lagrangian of the form \cite{nflation}:
\be
{\cal L}= \sum_{i=1}^N\left[\frac{1}{2} \, (\partial a_i)^2-\Lambda^4\left(1-\cos\left(\frac{a_i}{f_{a_i}}\right)\right)\right].
\label{Lnaxion}
\ee
Even though each single axion field may not be able to give rise to (natural) inflation for decay constants $f_{a_i}\ll M_P$, the ensemble of axion fields could reproduce similar features of a large field inflationary scenario. In fact, the predictions for the
cosmological observables are $0.93\leq n_s\leq 0.95$ and $r\leq 10^{-3}$ \cite{seery}. There is also a strong correlation between values of $r$ and the non-Gaussianity parameter $f_{\rm NL}$ with larger values of $5\leq f_{\rm NL}\leq 20$ corresponding to smaller values of $r\leq 5\cdot 10^{-4}$ \cite{seery}.

This is a string inspired scenario because of the presence of several ($N\sim 10^3$) axionic fields \cite{Grimm}.
However, in any concrete realisation of this model it is very difficult
to fix all the non-axionic moduli at an energy scale larger than the axionic potential.
Moreover, the Lagrangian (\ref{Lnaxion}) in general involves also cross-terms which
cannot be neglected and substantially complicate the inflationary dynamics challenging the identification of a single direction for collective motion \cite{Easther}.

\item{} {\it Axion monodromy}

This model \cite{MONODROMY} was designed with the goal of achieving a measurably large $r$. It has several realisations in terms of twisted tori or wrapped NS5 branes with the simplest potential of the form:
\be
V_{\rm AM}\sim \mu^3 \phi+\Lambda^4\cos\left(\frac{\phi}{f}\right)\,.
\label{VAM}
\ee
The key-ingredient to obtain a trans-Planckian field range for the inflaton is the
monodromy introduced by wrapped branes which `unwrap' the compact axionic direction.
The predictions for the cosmological observables are $0.97\leq n_s\leq 0.98$ and
observable tensor modes of order $0.04\leq r \leq 0.07$.
Ripples in the power-spectrum \cite{Flauger} and resonant non-Gaussianities \cite{Hannestad}
can be generated by the cosine modulation in (\ref{VAM}).\footnote{See also \cite{MonExp} for recent constraints on
monodromy inflation from WMAP9 data.}
In this case, warping is needed to render the non-axionic moduli heavy
in order to prevent a possible destabilisation of the inflationary potential.
Moreover, dangerous back-reaction issues have been discussed in \cite{MonodromyBR}.
\end{enumerate}

\subsubsection{Inflation using moduli}

Let us now turn to the discussion of models where the inflaton is a modulus of the extra dimensions; in particular the volume of a divisor of the internal Calabi-Yau space. Among the motivations for these models is the progress they can allow with the $\eta$-problem, partly due to the exponential form of the single-particle potentials to which they lead \cite{BBbarModuli,KMI,FIBRE,PolyInf}:
\be
V \simeq V_0 \left(1 - \beta k e^{-k\phi}  \right)\,.
\label{ModInfl}
\ee
This expression represents the approximation of the potential in the inflationary region where $\beta$ and $k$ are positive constants, with $\beta \simeq \mc{O}(1)$ while $k$ can be either $\mc{O}(1)$ or very large depending on the details of the model.

Exponential potentials such as this help with the $\eta$-problem because slow-roll only requires $\phi$ be sufficiently large
(sometimes trans-Planckian\footnote{Because the potential does not grow when $\phi$ is this large, even having trans-Planckian $\phi$ does not necessarily also produce observable tensor modes in these models.}), as is clear from the size of slow-roll parameters that follow from (\ref{ModInfl}):
\be
 \eta \simeq -\beta k^3 e^{-k\phi} \qquad\text{and}\qquad \epsilon \simeq \frac{\eta^2}{2k^2} \,.
\label{SRParam}
\ee
The point in these models is that moving to larger $\phi$ simply moves one further into the domain of validity of the low-energy theory.
This is because the canonically normalised modulus $\phi$ is often related to an extra-dimensional physical size, $R$, by a relation like $(\ell_s/R)^p \simeq e^{- k \phi}$, where $p$ is positive and $\ell_s$ is a length of order the string scale \cite{BBbarModuli}.

The hard part in these models is to obtain an approximately constant potential ({\em i.e.} the first term in eq.~\pref{ModInfl}), but this has proven to arise very naturally within large-volume string models \cite{CQreview}. This is because of an important feature all large-volume models share. Large-volume models systematically develop the low-energy field theory in powers of string coupling and the inverse size of the various extra-dimensional cycles that arise within Calabi-Yau geometries. The important feature for inflationary applications is that only a few moduli appear in the low-energy K\"ahler potential at the lowest orders of this expansion. The leading contribution,
\be \label{Ktree}
 K_{\rm tree} = -2\ln\vo \,,
\ee
depends only on the total extra-dimensional volume, $\vo$, and although in principle the leading $\alpha'$ \cite{BBHL} and string loop \cite{Kgs} corrections could have depended on all of the K\"ahler moduli they do not do so, due to the `extended no-scale structure' these theories enjoy \cite{ExtNoScale}. The other moduli eventually do appear at higher orders, but do so suppressed by additional powers of the small parameters that govern the expansions, and so any modulus orthogonal to $\vo$ has a comparatively flat potential, making it a natural inflaton candidate.

Another virtue of large-volume models is the protection they provide from `Planck slop': the contributions to the potential coming from dimension-6 Planck-suppressed operators within the low-energy theory. For instance eq.~\pref{Ktree} ensures that those arising within supergravity by expanding the prefactor $e^K$ of the F-term scalar potential are suppressed by $1/\vo^2$, and so depend at leading order just on the Calabi-Yau volume but not on the inflaton field. String loop corrections to $K$ can induce these higher-order operators but again, due to the extended no-scale structure, they tend to yield negligible contributions to the slow-roll parameter $\eta$.\footnote{See however \cite{grimm} who estimates an $\alpha'^2$ correction that could come with large coefficients and be suppressed by a factor of order $g_s^{3/2}{\cal V}^{-1/3}$. This is an interesting correction that could be relevant for models with only marginally large volumes and not too small values of $g_s$, and needs to be calculated model by model.}

The observational consequences of these models follow directly from the exponential potential\footnote{These consequences also resemble other models that share this potential, such as $R^2$ inflation \cite{RsqI} and Higgs inflation \cite{HI}.} (\ref{ModInfl}), which implies $\eta < 0$ and $\epsilon$ is at most of order the higher-order slow-roll parameters like $\xi \simeq 2 k^2 \epsilon \simeq \eta^2$. This leads to predictions for $n_s < 1$ and $r$ that agree extremely well with the data:
\be
r\simeq \frac{2}{k^2}\left(n_s-1\right)^2\qquad\text{and}\qquad \alpha_s \simeq -\frac 12 \left(n_s-1\right)^2\,.
\ee
Using the central value of the Planck result for the spectral index $n_s \simeq 0.96$, these become:
\be
r\simeq \frac{0.0032}{k^2}\qquad\text{and}\qquad \alpha_s \simeq -0.0008\,,
\label{rANDas}
\ee
giving a tensor-to-scalar ratio which depends on the value of the parameter $k$ and a $k$-independent running of the spectral index that agrees well with the Planck result $\alpha_s=- 0.013\pm 0.009$ at $68\%$ CL \cite{PlanckInf}.

Within this framework three successful inflationary models have been developed so far, which differ in the values predicted for $k$:
\begin{enumerate}
\item{} {\it K\"ahler Moduli Inflation}

In this model \cite{KMI} the inflaton is the volume of a blow-up mode and its potential is developed by non-perturbative effects.\footnote{In retrospect a more appropriate name might have been `blow-up inflation' since K\"ahler moduli also include fibre moduli which are also inflaton candidates as seen below.}
This is the first inflationary model that emerged from the large-volume scenario \cite{LVS} of modulus stabilisation. The typical case corresponds to what is known as a Swiss cheese Calabi-Yau compactification with one large K\"ahler modulus and several blow-up modes corresponding to the sizes of the four-cycles that can be collapsed to zero size keeping the overall volume stable.

The canonical inflaton potential predicted is very similar to (\ref{ModInfl}) with a very large parameter\footnote{Notice that in the original model proposed in \cite{KMI}, $k\propto \vo^{2/3}$. The difference comes from the fact that here we are expanding the inflaton potential around its minimum.} $k\propto \vo^{1/2}\ln \vo$. Because a successful amplitude of scalar perturbations requires an internal volume of order $\vo\simeq 10^6$ in string units, $k \gg 1$ and so the inflationary scale is around $10^{13}$ GeV and the tensor-to-scalar ratio is $r\simeq 10^{-10}$. The relation between the spectral index and the number of $e$-folds is $n_s\simeq 1-\frac{2}{N_e}$ while the running spectral index is $\alpha_s \simeq -\frac{4}{N_e^2}$.
Hence for $50\leq N_e \leq 60$, one obtains the predictions $0.96\leq n_s\leq 0.967$ and $-0.0016\leq\alpha_s\leq-0.001$.
The simplest models do not give detectable non-gaussianities but extensions of them can incorporate the curvaton mechanism which allows for $f_{\rm NL}^{\rm local}\sim \mc{O} (10)$ \cite{StrNG1}.
A shortcoming of this model on the theoretical side is that string loop effects tend to spoil the flatness of the inflationary potential in the region close to the minimum where non-perturbative effects give rise to slow-roll inflation. Thus in order to obtain enough $e$-foldings of inflation one has to suppress the $g_s$ effects by a suitable tuning of their coefficients.

\item{} {\it Fibre Inflation}

In this model \cite{FIBRE} the inflaton is the volume of a K3 or $T^4$ divisor fibred over a $\mathbb{P}^1$ base.
The inflationary potential is developed by string loop corrections which, contrary to `K\"ahler moduli inflation',
can give rise to a sufficiently long period of inflation without the need of tuning any underlying parameter.
The reason is the natural hierarchy of masses amongst the moduli in string constructions which naturally lead to a potential
of the form (\ref{ModInfl}) with $k = 1/\sqrt{3}$.

This is a large-field inflation model, whose exponential form naturally imposes a relation between the slow-roll parameters: $\epsilon \simeq \frac 32 \eta^2$, implying the prediction $r \simeq 6(n_s-1)^2$. Even though small, this is potentially observable (although most probably not by the Planck polarization analysis). For $50\leq N_e\leq 60$ we have $0.965\leq n_s \leq 0.97$ and $0.005\leq r\leq 0.007$. The inflationary scale is $M_{\rm inf}\simeq 5\times 10^{15} $ GeV. The simplest models give unobservable values of $f_{\rm NL}$ but extensions of these models give rise to non-Gaussianities from modulated reheating with $f_{\rm NL}$ of order `a few' \cite{StrNG1}.

\item{}
{\it Poly-instanton Inflation}

In this model \cite{PolyInf} the inflaton is again the volume of a K3 or $T^4$ fibre over a $\mathbb{P}^1$
base but its potential is now generated by non-perturbative effects, namely by poly-instanton
corrections to the superpotential. Contrary to `K\"ahler moduli inflation', in this case,
string loop effects are harmless since inflation takes place in a region closer to the minimum of
the potential were $g_s$ effects can be shown to be still negligible for natural values of the underlying parameters.
The inflationary potential look like (\ref{ModInfl}) with $k\propto \ln \vo$. If we now look at the $k$ dependence of the slow-roll parameters (\ref{SRParam}),
we notice that in order to have $|\eta|\ll 1$ larger values of $k$ require larger values of $\phi$.
This is the simple reason why in this case the inflationary region is closer to the minimum than in `K\"ahler moduli inflation'
where $k\propto \vo^{1/2}\ln\vo$.

The predictions of this model for 54 $e$-foldings of inflation (as required by a reheating temperature of order $T_{\rm rh}\simeq 10^6$ GeV)
are: a Calabi-Yau volume of order $\vo \simeq 10^3$ in string units, a high inflationary scale, $M_{\rm inf}\simeq 10^{15}$ GeV,
a small tensor-to-scalar ratio, $r\simeq 10^{-5}$, and a spectral index which is perfectly in agreement with observations:
$n_s \simeq 0.96$.
\end{enumerate}

\subsection{The $n_s$ vs $r$ plane}

\begin{figure}[tbp]
\includegraphics[width=.58\textwidth]{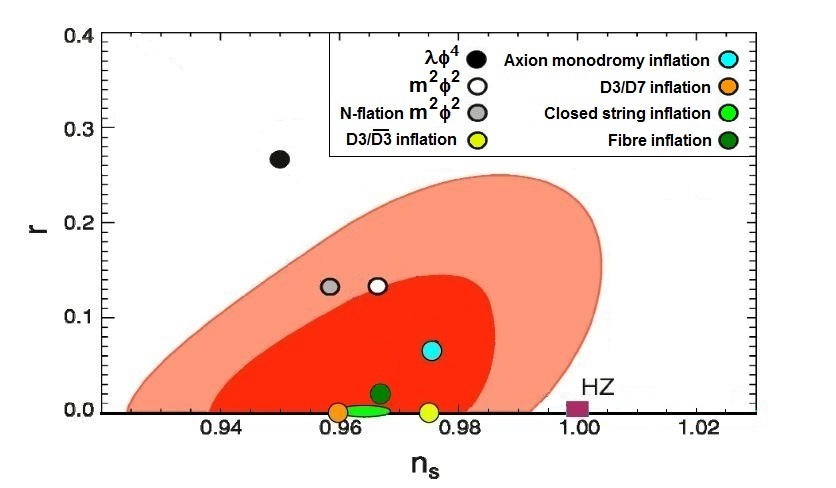}
\hfill
\includegraphics[width=.45\textwidth,origin=c]{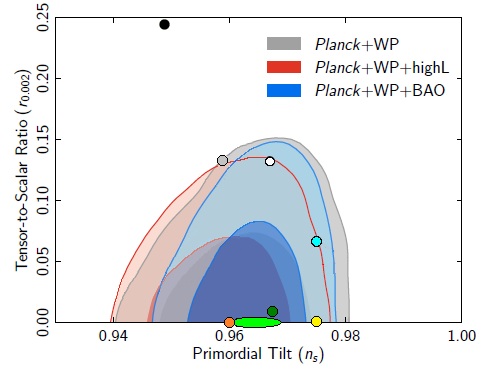}
\caption{\label{nsvrfig} Left panel: A comparison of WMAP constraints in the $n_s$-$r$ plane with several string models for $N_e\simeq 60$, taken from an ICHEP 2008 summary talk \cite{ICHEP2008}. Right panel: The same comparison superimposed on the Planck constraints taken from \cite{Planck}, with `D3/$\overline{{\rm D3}}$ inflation' (yellow oval); `D3/D7 inflation' (orange oval); `closed-string inflation' (light green oval); `Fibre inflation' (dark green oval) and `Axion monodromy inflation' (cyan oval).}
\end{figure}

Predictions of the scalar spectral tilt, $n_s$, and the primordial scalar-to-tensor ratio, $r$, are the bread and butter of comparisons of inflationary models with observations. As functions of the slow-roll parameters, $\epsilon$ and $\eta$, single-field models predict these to be \cite{nsr}:
\be \label{nsrvsslowroll}
 n_s - 1 = 2\eta-6 \epsilon\qquad \hbox{and} \qquad
 r = 16 \,\epsilon \,.
\ee
The experimental constraints on these coming from WMAP \cite{WMAP} and Planck \cite{PlanckInf} are respectively given on the left- and right-hand panels of Fig.~\ref{nsvrfig}, which both show an error ellipse that encircles a central point near $n_s \simeq 0.96$ with $r$ consistent with zero.

Also shown in Fig.~\ref{nsvrfig} are the predictions of several inflationary models. Most of these are provided for comparison purposes by the observers themselves, but to these have been added a collection of string-inflationary predictions as well. All of these predictions are pre-2013, since the left-hand panel is taken directly from a 2008 summary \cite{ICHEP2008} given at the ICHEP meeting. We have transferred these by eye in the right-hand panel to the Planck results for ease of comparison. For convenience we also collect the predictions of the above models for $50\leq N_e\leq 60$ in the table below.

\be
\begin{array}{|c|c|c|}
\hline {\rm String \, Scenario} & n_s &  r  \tabularnewline \hline \hline
       {\rm D3/{\overline{\rm D3}} \, Inflation}  &  0.966\leq n_s\leq 0.972  &   r\leq 10^{-5}  \tabularnewline\hline
       {\rm Inflection\,  Point\,  Inflation}  &  0.92\leq n_s\leq 0.93  &   r\leq 10^{-6} \tabularnewline\hline
       {\rm DBI\, Inflation}  &  0.93\leq n_s\leq 0.93  &  r \leq 10^{-7}   \tabularnewline\hline
       {\rm Wilson\,  Line\,  Inflation}  &  0.96\leq n_s\leq 0.97  &  r\leq 10^{-10}   \tabularnewline\hline
       {\rm D3/D7 \, Inflation}  &  0.95\leq n_s\leq 0.97  &  10^{-12}\leq r\leq 10^{-5} \tabularnewline\hline
       {\rm Racetrack\,  Inflation}  &  0.95\leq n_s\leq 0.96  &  r\leq 10^{-8}   \tabularnewline\hline
       {\rm N-flation}  &  0.93\leq n_s\leq 0.95  &  r\leq 10^{-3}   \tabularnewline\hline
       {\rm Axion\, Monodromy}  &  0.97\leq n_s\leq 0.98  &  0.04\leq r\leq 0.07   \tabularnewline\hline
       {\rm Kahler\,  Moduli\,  Inflation} &  0.96\leq n_s\leq 0.967 &  r\leq 10^{-10}  \tabularnewline\hline
       {\rm Fibre\,  Inflation}  &  0.965\leq n_s\leq 0.97 &  0.0057\leq r\leq 0.007 \tabularnewline\hline
       {\rm Poly-instanton\, Inflation}  &  0.95\leq n_s\leq 0.97  &  r\leq 10^{-5}  \tabularnewline\hline
\end{array} \nonumber
\label{table}\,,
\ee

\vspace{1cm}

Of the models depicted, `D3/$\overline{{\rm D3}}$ inflation' \cite{KKLMMT} represents the predictions of the first bona-fide string implementation of brane-antibrane inflation \cite{BBbarInf, Dvali:2001fw}, including modulus stabilisation. The orange oval marked `D3/D7 inflation' \cite{HAACK}
and the light green oval marked `closed string inflation' represent the predictions of a broad class of models \cite{Racetrack, KMI, PolyInf, StrNG1, StrNG2, StrNG3} which differ somewhat in their predictions for $\eta$, but all find $\epsilon$ too small to show $r$ non-zero on the plot. Notice that similar predictions are obtained in models where inflation is obtained from wrapped D-branes \cite{wrappedbranes}, inflection points \cite{delicate}, Wilson lines \cite{wilsonline} or non-canonical kinetic terms \cite{DBI}. All of these models describe the observed fluctuations very well, and much better than simple single-field $\phi^2$ models.

Apart from `N-flation' \cite{nflation} which suffers from the control issues mentioned above, only two of the string models, `Axion monodromy inflation' \cite{MONODROMY} and `Fibre inflation' \cite{FIBRE}, predict $r$ large enough to be visible on the plot. These two were specifically designed for the purpose of obtaining large $r$, since it had been remarked that small $r$ appeared to be generic to string-inflationary models. They both score reasonably well for the $\eta$-problem, but both have also been criticized. Ref.~\cite{MonodromyBR} argues that the lack of supersymmetry in the models of ref.~\cite{MONODROMY} can make it more difficult to control the corrections to leading predictions, with potentially significant back-reaction effects. The `Fibre inflation' model builds on the hierarchy of masses that loops and higher-derivative corrections introduce into the low-energy potential, but in the absence of their explicit calculation must use an educated guess for their detailed shape.

Although present models cannot claim to explore all of string parameter space, it is striking how unanimously they predict small $r$, and how well their predictions agree with observations. Is there a reason for this agreement? Possibly, as we now see.

\subsection{Future prospects for measuring $r$}

Forecasting the expected size of primordial tensor perturbations is particularly useful now given that observations are likely to become significantly more sensitive to $r$ in the near future. What might these observations expect to find? Time for theorists to nail their colours to the mast.

As eq.~\pref{nsrvsslowroll} shows, a theory's position in the $n_s$ - $r$ plane is dictated by the two slow-roll parameters, $\epsilon$ and $\eta$. One combination of these two parameters is determined from the value of $n_s - 1 \simeq 0.04$ inferred from observations. Opinions about the likelihood of $r$ being observable then come down to opinions about how big $\epsilon$ might be. Two points of view towards what should be expected are widely touted. These are:
\begin{itemize}
\item {\em Flat prior:} One point of view argues that in the absence of other information the two small quantities $\epsilon$ and $\eta$ should be expected to be similar in size, so if inflation is true then tensor modes should soon be observed \cite{Slava,TensorsLikely}.
\item {\em Flat log prior:} A second point of view starts from the observation that the size of primordial tensor perturbations is purely set by the size of the dominant energy density during inflation, and this could be anywhere between 100 GeV and $10^{15}$ GeV. Since only the top end of this range produces an observable value for $r$, this point of view argues that since we have no intrinsic reason to prefer any scale over any other we should have no preference for observable or unobservable $r$.
\end{itemize}
To these we would add one more, string-motivated point of view:
\begin{itemize}
\item {\em Trans-Planckian fields:} In single-field models the value of $\epsilon$ can be related to the distance traveled by the inflaton between horizon exit and inflation's end \cite{LythBound}, with an observably large value of $r$ corresponding to trans-Planckian field motion:\footnote{See ref.~\cite{Arthur,LythUnbound} for explorations of loopholes to this bound.} $\Delta \phi \gsim M_p$. Whether such a large field excursion is inconsistent with the low-energy limit that is implicit in using any gravitational theory depends on the energy cost associated with having large fields, but this is difficult to judge without having a controlled understanding of physics near the Planck scale. Here string models can help since they provide a framework for assessing how hard it is to obtain these Planckian field excursions. It is the difficulty in doing so that has made obtaining large $r$ difficult in string models, and this difficulty has led some to seek no-go theorems of various types \cite{QGconstraintsInf}. Short of such a no-go, the phenomenological observation that the majority of known string models do not predict such excursions suggests the point of view that we should {\em expect} $r$ to be too small to be visible. This statement goes in both ways since if tensor modes are discovered this will single out a very small class of string inflationary models.
\end{itemize}

Notice that the flat prior assumes a generic potential while string theory tends to produce inflationary potentials
which are usually not generic. A good counterexample to the flat prior described above, is given by
K\"ahler moduli, Fibre and Poly-instanton inflation where the scalar potential takes the particular shape (\ref{ModInfl})
which leads to the \emph{natural} prediction $\epsilon\ll|\eta|\ll 1$ in perfect agreement with present Planck data.

\subsection{Non-Gaussianity and emergent single-field models}

Perhaps the biggest disappointment in the Planck results was the absence of evidence for primordial non-Gaussianity. Discovery of primordial non-Gaussianity would have opened up an entirely new class of physically informative functions to be measured in the sky. Single-field slow-roll inflationary models robustly predict primordial non-Gaussianity to be too small to observe, in agreement with what is observed. What would string inflationary models expect for this?

There are two mechanisms for generating non-Gaussianity that have been proposed within string inflationary models. One way is by modifying the single-field dynamics of the inflaton, such as by having its kinetic energy be described by a DBI Lagrangian \cite{DBI}. These modifications tend to produce non-Gaussianity at horizon exit, producing `equilateral' bi-spectrum distributions. In order to be in agreement with recent Planck data for $f_{\rm NL}^{\rm eq}$ within $2\sigma$, the parameter $\beta$ introduced in eq.~\pref{DBIpots} has to be less than unity \cite{KKLMMTPlanck}. Since $\beta > 0.1$ observations constrain it to lie in the small window $0.1<\beta<\mathcal{O}(1)$. If the inflaton is described by brane motion, this motion seems restricted to lie in an unwarped region of the extra dimensions.

Another way non-Gaussianities can arise in string models is through multiple-field dynamics. The class of 4D models that one obtains at low-energies in string theory usually involves multiple fields, and it is within the scalar potential for this low-energy theory that an inflating classical trajectory is found. Although it is usually not generic to have any of these scalars be lighter than the Hubble scale, once this is arranged for one of them it also tends to be true for several of the others as well. Non-Gaussianities can then be obtained using mechanisms identified in multi-field inflationary models, such as the curvaton \cite{Curvaton} or modulation \cite{Modulation} mechanisms, and string-inflationary models have been found that realise both of these \cite{StrNG1,StrNG2}.\footnote{For other stringy realisations of multi-field dynamics which can give rise to large non-Gaussianities see \cite{StrNG3}.} These mechanisms tend to produce a bi-spectrum of the `local' form.

These examples show that non-Gaussianity can, but need not, be produced by string models. But one might ask whether or not non-Gaussianity should be regarded as generic. Because all string models infer the existence of inflation within the low-energy 4D theory, we are limited by the state of the art to answer this question within that context: what is generic of inflation in low-energy 4D models? For models with multiple light fields one might expect non-Gaussianity and isocurvature fluctuations to be the rule rather than the exception. But although isocurvature modes indeed can be generated during inflation, they need not be. And even if they are, generically they need not survive the epoch between inflation and later cosmology to imprint themselves observably onto the CMB.

Multi-field models often do not generate isocurvature perturbations because their observable cosmological consequences are very often well-described for all practical purposes by an effective single-field model. This always happens when the motion is along a trough, with all but one of the fields having masses larger than the Hubble scale \cite{Trough}, but it also often happens when many fields are light and so in motion during inflation, but when the motion does not deviate from a target-space geodesic near horizon exit. Single-field models are often `emergent' in this sense, and realising that this is true can sometimes allow more robust conclusions to be inferred \cite{RobustSF} than would be possible directly working with the full multi-field potential. This phenomenon has recently also been studied in more detail within random-field models motivated by string inflation \cite{Random}, allowing a more systematic search for inflationary solutions within complicated potentials.

This observation also provides a partial explanation of why inflationary models do so well describing the Planck data, despite the models consisting of complicated dynamics of multiple degrees of freedom and the data's preference for a simple single-field description. Although the underlying string dynamics is complicated, at crucial epochs like horizon exit it is often well-described by simple single-field physics and so gets the observations right. From this we might draw two general lessons: observational evidence for single-field models is not good evidence against multiple-field models; and we should generically expect single-field observations to describe the data well.

\section{Late-epoch relics}

It has long been known that the production of late-time relic particles and fields can also provide constraints on fundamental theories like string models \cite{ModulusProblem}. In fact, hidden sectors are ubiquitous in string compactifications,
and every time reheating takes place via the decay of a gauge singlet (like a gravitationally coupled modulus, for example),
a priori there is no fundamental reason why hidden sector particles should not be produced \cite{StringReheating}.
In turn, this might induce problems associated with the over-production of either dark matter or dark radiation.

This has been emphasized again recently in the context of large-volume models which can often leave relic forms of dark radiation in the late-time universe \cite{DRadLV}. To the extent that there is evidence from cosmology that the effective number of neutrinos $N_{eff}$ exceeds the Standard Model value $N_{eff}=3.04$ the possibility arises of using light relics as observable features of the hidden sectors. This observation has recently been made as a broader constraint for string compactifications \cite{DRadGeneral} in general. While we agree with the general spirit of these kinds of bounds, we would emphasize that they can also be somewhat model-dependent and contingent on what else may be going on in the universe at the time of interest. For instance, the following parts of parameter space could evade the general bounds of ref.~\cite{DRadGeneral}:
\begin{itemize}
\item The relic dark radiation could be produced before the QCD phase transition, because in this case the subsequent reheating of the observable sector can easily dilute a few species of new dark radiation.

\item The last scalar which decays when it dominates the energy density of the universe might not be a Standard Model gauge singlet, in which case the branching ratio into hidden sector degrees of freedom could be negligible and any dark radiation produced before this last decay could be diluted away.

\item A second lower-energy period of inflation (possibly driven by thermal effects \cite{TI}) could dilute any unwanted relic particles (see, however, \cite{2step} for potential hazards of scenarios with two inflationary periods).

\item The production rate of light hidden sector degrees of freedom might be highly suppressed relative to that of visible sector particles due to the details of the realisation of the Standard Model in a given compactification. For example, in the model described in \cite{DRadLV} the production of light axion dark radiation could be suppressed if the axions arise as open string states, or the visible sector is not sequestered, and so on.

\item Putative light candidate dark radiation particles might actually be heavier once higher-order effects are included. For instance axions could become heavy from non-perturbative effects, or be eaten up by anomalous $U(1)$s. Notice that this might even be forced by consistency conditions, like Freed-Witten anomaly cancellation. This is actually often the case in the explicit compact Calabi-Yau constructions with both chiral matter and stabilised moduli constructed recently \cite{ExplicitConstructions}.
\end{itemize}
Dark radiation overproduction may yet provide an important constraint on string model building into the future, but the power of this constraint can depend both on formal aspects related to the consistency of a given compactification and on the precise details of the model's predicted cosmological evolution.

\section{Summary}

We close with a brief summary of our summary. Today we have the luxury of access to unprecedented precision in our knowledge of many features of the early universe, most notably the recent Planck observations of the CMB and the associated inference of the properties of primordial fluctuations. These observations remain well-described by simple single-field inflationary models, although some of the past favourites like $\phi^2$ or $\phi^4$ models are becoming less favoured.

Can (or should) complicated string model building survive in the face of all this simplicity? We argue here that it can (and should). String models are complicated because they try to control approximations about which model-builders in cosmology are often insouciant. The successes of inflationary models hinge on the interplay between quantum effects ({\em e.g.} the source of primordial fluctuations) with classical gravity. Quantum fluctuations are by design large enough to be observable, and the non-trivial interplay between quantum and gravitational dynamics pushes the envelope of the quantum-gravity frontier. The same is even more true for models whose formulation intrinsically requires strong-gravity physics, such as a universal bounce. It is incumbent on us to embed such models into a sensible theory of quantum gravity in order to properly quantify their theoretical uncertainties. At present string theory is the only such a framework available that is sufficiently well-formulated to allow this to be done with some precision.

Yet this complication does not appear also to mean a lack of predictive power. The models to date predict values of $n_s$ and $r$ that cluster very well within the preferred region of the Planck 2013 data. They largely do so because their predictions for $r$ tend to be very small, likely due to the difficulty in obtaining controlled trans-Planckian motion in field space. A few models have been constructed to produce larger values for $r$, but even these were not able to be large enough to be in serious tension with the Planck data. Although parameter space will continue to be explored, on present evidence it seems difficult to obtain trans-Planckian rolls within the one framework (string theory) within which the possibility can be explored in a controlled way. The difficulty in doing so suggests we are unlikely to see primordial tensor modes as observational results are improved. Put more optimistically, if they are seen it would rule out most string-inflation models --- perhaps the most informative result for which one could hope.

How can complicated models be consistent with observations that favour simplicity? Because evidence for simple single-field models need not be evidence against more complicated multiple-field dynamics. As we explore the complicated multi-field dynamics of more and more string examples we see that the key cosmological observables are often well-described by simple emergent effective single-field dynamics. Single-field models do well because the multiple-field dynamics happens not to do much interesting during crucial epochs like horizon exit. Indeed it is the identification of robust features such as this that is part of the point of exploring the more complicated theories that arise as the low-energy limit of string constructions.

Bring on the new observations!

\subsection*{Acknowledgements}

We thank Daniel Baumann, Latham Boyle, Shanta de Alwis, Liam McAllister, Hiranya Peiris and Kendrick Smith for useful discussions. CB wishes to thank the Abdus Salam International Centre for Theoretical Physics (ICTP) and the Bethe Center for Theoretical Physics (BCTP) for their support, and for providing the pleasant yet stimulating environments within which part of this work was done. He also thanks the organisers of the Planck 2013 conference where some of the results were presented. Releasing the paper after the conference avoided any ambiguities with the title. Research was supported in part by funds from the Natural Sciences and Engineering Research Council (NSERC) of Canada. Research at the Perimeter Institute is supported in part by the Government of Canada through Industry Canada, and by the Province of Ontario through the Ministry of Research and Information (MRI).

\end{document}